\documentclass[fleqn,twoside]{article}
\usepackage{espcrc2}

\usepackage{graphicx}
\usepackage[figuresright]{rotating}

\newcommand{\AmS}{{\protect\the\textfont2
  A\kern-.1667em\lower.5ex\hbox{M}\kern-.125emS}}

\title{Non-perturbative propagators and dimension 2 condensate in Yang-Mills theory%
\thanks{Supported by
Sumitomo Foundations, Grant-in-Aid for Scientific Research  (B)13135203 
from MEXT,  and  (C)14540243  from JSPS.}
}

\author{Takeharu Murakami\address[CUGS]{Graduate School of Science and Technology, 
           Chiba University, Chiba 263-8522, Japan}\thanks{Speaker at the conference.},
        Kei-Ichi Kondo\address[CUDP]{Department of Physics, Faculty of Science, 
           Chiba University, Chiba 263-8522, Japan},
        Toru Shinohara\addressmark[CUGS]
  and   Akihiro Shibata\address{National Laboratory for High Energy Physics (KEK), Tsukuba, Ibaraki 305, Japan}
}

\usepackage{amsmath}

\def\green#1{#1}

\def\red#1{#1}
\def\blue#1{#1}

\def\cyan#1{}

\begin{document}

\begin{abstract}
We have found ultraviolet asymptotic slutions of the Schwinger-Dyson equation 
for the gluon and ghost propagators which have simultaneously
 the perturbative logarithmic correction and the non-perturbative $1/p^2$ power correction. 
By including the perturbative corrections, the power correction reproduces exactly the 
leading OPE result suggesting the existence of dimension two condensate. 
\end{abstract}

\maketitle


\section{ Introduction }
Various condensates are related to non-perturbative properties of QCD. 
The importance of the condensates such as $\langle FF\rangle$ and  $\langle \bar{q}q\rangle$ are well-known.

 Though the dimension two operators such as mass terms
are not BRST invariant in Yang-Mills (YM) theory,
 it was pointed out recently that these operators 
contain a gauge invariant physical part 
and that a special combination of them can be (on-shell) BRST invariant
\cite{Zakharov}
\cite{Kondo01}
\cite{Boucaud}.

It was argued that the minimum of $ A^2$ 
along the gauge orbit can have a definite physical meaning, 
and how to define the physical part non-perturbatively \cite{Zakharov}.


There is a BRST-invariant composite operator of mass dimension 2 \cite{Kondo01} 
as a linear combination of $A^2$ and quadratic ghost, averaged in spacetime:
{\footnotesize
\begin{equation}
   \mathcal{O}  
  = \Omega^{-1} \int d^4x \ \text{tr}\left[ {1 \over 2} A^\mu(x) A_\mu(x) + \lambda i \bar{C}(x) C(x) \right].
\end{equation}
}
In the Lorentz gauge, it reduces to
 the same form as the known Curci-Ferrari mass term.
But we do not include it in the Lagrangian. 
Therefore our theory is usual YM. 

It is easily seen that this operator is on-shell BRST invariant: 
$\delta_{BRST} \mathcal{O}=0$.
The $A^2$ part should be divided into the physical (transverse) part 
and other unphysical (longitudinal and scalar) part.
When we take the vacuum expectation value, 
the $\bar{C}C$ part precisely cancels this unphysical part.
Thus the operator has a gauge invariant expectation value,
though the remaining physical part is nonlocal. 
Especially in the Landau gauge  $\lambda \rightarrow 0$, 
$\langle {\cal O} \rangle \rightarrow \langle {1 \over 2} A^\mu(0) A_\mu(0) \rangle$.

Since these operators are not BRST invariant as local polynomials, 
they do not appear in 
OPE of usual gauge invariant quantities. But 
they may appear in OPE of 
BRST non-invariant quantities such as \blue{propagators}. 


 In fact, this operator in the Landau gauge has been estimated in various methods. 
Boucaud et al.\cite{Boucaud} have simulated the lattice propagator and used OPE fit 
to obtain $\langle A^2\rangle \simeq (1.4GeV)^2 $. 
Verschelde et al.\ \cite{Verschelde} discussed the effective potential of $A^2$. 


Here we take the Schwinger-Dyson (SD) equation approach. 
Recent investigations \cite{SHA}\cite{AB} of
 Euclidean pure SU$(N_c)$ Yang-Mills theory in the Lorentz gauge show
{\small
\begin{equation}
  \lim_{p^2 \rightarrow 0} p^2 D_T(p^2)  = 0 ,\ \ 
  \lim_{p^2 \rightarrow 0} p^2 G_{gh}(p^2)  = \infty .
\end{equation}
}
The transverse gluon propagator vanishes, 
while the FP ghost propagator is enhanced in the infrared limit $p^2 \rightarrow 0$. 
That is,
 the ghost propagator behaves more singularly than the free propagator in low energy region. 
These results mean the IR ghost dominance. 
This is consistent with 
 the well-known Gribov prediction and the confinement criterion due to Kugo and Ojima.

But these results are strongly dependent on approximations in the calculation. 
Simple bare vertex approximation fails to reproduce 1-loop perturbative result.
This situation is very different from the case of fermion SD equation. 

 In addition to the approximation ambiguity, 
we pay attention to the \red{$1/p^2$ corrections} in the UV propagators 
which may come from OPE 
($\langle A^2 \rangle $ condensate)
or renormalon or some other non-perturbative effects.

\section{
SD equations of Yang-Mills theory in the Landau gauge
}

We consider the SD equation for the gluon propagator $D$ and ghost propagator $\Delta$ 
in pure YM theory in the Landau gauge. 
For the gluon propagator $D$, we adopt the Brown-Penington projection 
$
  R_{\mu\nu}(p) := \delta_{\mu\nu} - 4{p_\mu p_\nu \over p^2} 
$
to remove a tadpole graph.

Next,  we neglect all 2-loop diagrams 
and adopt Higashijima-Miransky approximation
to the internal propagators. 

Thus the relevant SD equations become
\vspace*{-2mm}
\[
\begin{array}{c}
\includegraphics[width=55mm]{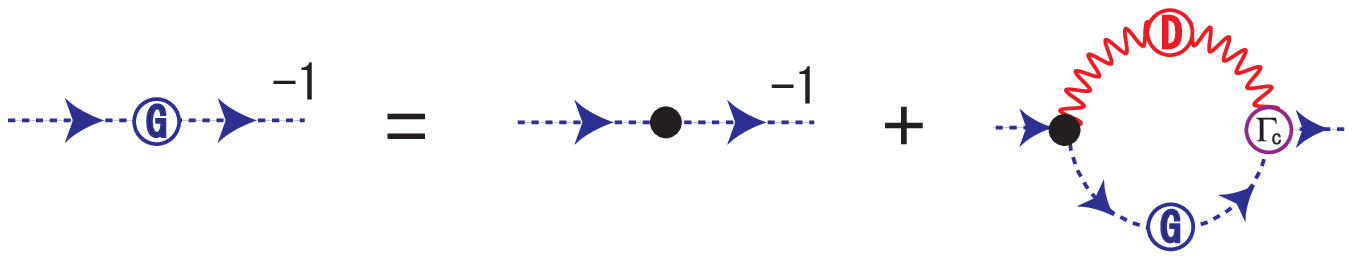}
\end{array}
\]
\vspace*{-3mm}
\[
\begin{array}{c}
\includegraphics[width=55mm]{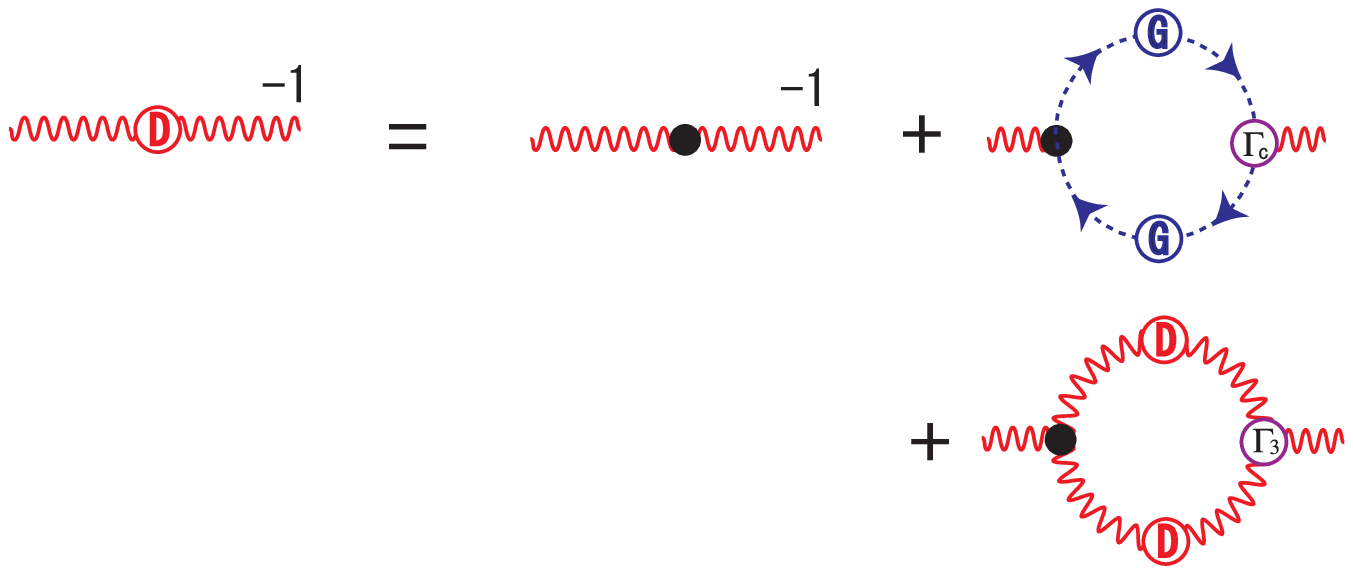}
\end{array}
\]
\vspace*{-8mm}


\section{Vertex corrections}
At this stage,
 in addition to propagators $D$ and $G$, the vertices $\Gamma_c$ and $\Gamma_3$ are unknown. 
In the Landau gauge, 
the ghost-gluon vertex $\Gamma_c$ does not have divergence. 
So our main attention directs to approximate $\Gamma_3$.

In the investigations \cite{SHA} \cite{AB} \cite{FAR} \cite{Bloch}, 
various ansatz are adopted to $\Gamma_3$. 
In \cite{SHA} \cite{AB}, the bare or the Slavnov-Taylor (ST) improved vertex is adopted. 
This procedure is similar to the case of fermion SD. 
In \cite{FAR} \cite{Bloch}, 
improved vertices are used to reproduce the UV 1-loop perturbative result correctly where 
 free parameters $a$ and $b$ are introduced in  \cite{FAR}. 
It does not change UV leading exponent. 
Bloch \cite{Bloch} used a similar ansatz 
derived from his own argument based on multiplicative renormalizability. 
But in their cases, infrared solutions become very delicate in the sense that 
whether ghost dominance property is realized or not depends on 
the choice of parameters or the effect of 2-loop diagrams.

\section{ UV asymptotic solutions of the SD equations}

The simple bare vertex approximation cannot reproduce even the 1-loop perturbative result, 
while the IR solutions are very sensitive to the choice of vertex correction. 
In order to get rid of this difficulty, 
we consider the higher logarithmic terms in the UV asymptotic solution. 
We are interested in the ``non-perturbative" power correction terms too.

To obtain the solutions, 
 we define the gluon form factor $F$ and the 
ghost one $G$ by multiplying $p^2$ to the propagators. 

For the UV asymptotic solutions,
 we adopt the new ansatz which has logarithmic (perturbative) powers 
and power (non-perturbative) corrections as \cite{Kondo02PL}
{\footnotesize
\begin{eqnarray}
  F(p^2) = A \blue{(\omega\ln p^2)^{\gamma}} \sum_{n=0}^{N} c_n (\frac{1}{\omega\ln p^2})^{n}   
+  \frac{(\ln p^2)^{\gamma+\gamma_1}}{p^2} a^{(1)}   ,
\nonumber\hspace*{-1mm}
\\
  G(p^2) = B \blue{(\omega\ln p^2)^{\delta}} \sum_{n=0}^{N} d_n (\frac{1}{\omega\ln p^2})^{n} 
+ \frac{(\ln p^2)^{\delta+\delta_1}}{p^2}  b^{(1)}.
\label{soln_ansatz}
\end{eqnarray}
}

We use the two parameter ansatz \cite{FAR} for the renormalized triple gluon vertex
{\footnotesize
\begin{eqnarray}
 Z_1 \Gamma_3(p,q) 
  := \frac{G(q^2)^{1-\frac{a}{\delta}-2a} G((p-q)^2)^{1-\frac{b}{\delta}-2b}}{F(q^2)^{1+a}F((p-q)^2)^{1+b}}.
\end{eqnarray}
}
Simply introducing parameters makes a result more ambiguous. 
In our approach, new parameters will be automatically determined
by considering subleading solutions and OPE consistency.

We put these into SD equations, and get infinite series of algebraic equations
 with respect to
coefficients $A, B,$ $\omega, \gamma, \delta,$ $c_n, d_n,$ $ a^{(1)}, b^{(1)},$ $\gamma_1, \delta_1$ and parameters $a, b$.


\section{ Result}
We first find that 
without power corrections ($a^{(1)}=0$, $b^{(1)}=0$) 
the ansatz cannot satisfy the coupled SD equation for large $p^2$, 
even if we include the logarithmic corrections $c_n, d_n\not=0$.   

A self-consistent solution is obtained when
{\footnotesize
\begin{eqnarray}
 \blue{\gamma} &=& - 13/22 ,\ \delta = -  9/44 \\
\omega &=& \frac{11}{3}N_c \lambda \tilde{Z}_1 AB^2
= \frac{11}{3}N_c \lambda  = \beta_0 \ \ (\lambda:=\frac{g^2}{16\pi^2}) \nonumber
\end{eqnarray}
}
for arbitrary value of $a$ and $b$.

These reproduce leading exponents and beta function in agreement with 1-loop perturbation.

The remaining coefficients are determined successively 
except for one degrees of freedom $(a+b)$. 
{\footnotesize
\begin{eqnarray}
 (\red{a+b}) &=& -\frac{421}{126}+\frac{9152}{945}\red{d_1}, \nonumber\\
 \blue{\gamma+\gamma_1} &=& -1.28-\frac{9}{88}(a+b),\quad  \blue{\delta_1}=\gamma_1-1, \nonumber\\
b^{(1)} &=&\frac{9N_c}{8}\lambda\tilde{Z}_1 AB^2 \green{a^{(1)}},\nonumber\\
c_1&=&\frac{35}{24}N_c \lambda +\frac{26}{9}d_1,\nonumber\\
d_2&=&(4202070705N_c \lambda^3-958366464N_c \lambda^2 d_1\nonumber\\
&&+6683578880N_c \lambda d_1^2-39086260224 d_1^3)\nonumber\\
&&/(304128(645N_c \lambda-43648 d_1)),\nonumber\\
c_2&=& \cdots\nonumber\\
 \ \vdots 
\label{results}
\end{eqnarray}
}
Coefficients $A$ and $B$ (overall normalization)
will be determined if we fix the renormalization condition. 
$a^{(1)}$ and $b^{(1)}$ (coefficients of power correction) are not determined. 
Except for $d_1$ (or \red{$a+b$}), 
all the other coefficients and exponents
 are calculated by simple algebraic equations up to any finite orders. 
The logarithmic expansions seem to converge in UV region.


\section{ Comparison with OPE result}
The additional logarithmic exponent of the power correction 
calculated from OPE is 
$-(1-{\hat{\gamma}_0 \over \beta_0})=-35/44=-0.795$. 
Corresponding exponent from the SD solutions in bare vertex approximation is 
$\gamma+\gamma_1 =-0.935$ \cite{Kondo02PL}.

In the improved vertex case (\ref{results}), 
only one unknown parameter $(a+b)$ governs this exponent. 
We use the relation (\ref{results}) to determine the parameter.
When we set $a+b=-4.778$ (or $d_1 = -0.148$), 
the exponent $\gamma+\gamma_1$ becomes $-0.795$ 
which reproduces the leading OPE. 
Moreover, this value
for $a+b$ 
leads to the ghost dominance solution in the IR region.


\section{Conclusion}

We have found the asymptotic 
SD solutions of the gluon and ghost propagators in 
SU($N$) Yang-Mills in Landau gauge for the ansatz using power and logarithmic expansion.

``Non-perturbative" \blue{$O(1/p^2)$ power corrections} are necessary. 
This corresponds to dimension two vacuum condensates [1-3]. 

Contrary to the fermion SD case, 
the bare vertex approximation cannot reproduce the UV solution consistent with 
the UV perturbation and the UV OPE. 
We have found the consistent solution 
by considering the UV higher order corrections.


Using the same vertex ansatz (4) determined as above,
we have obtained also the IR solutions 
which have the same form as the power-law solutions [6]: 
$ F(p^2) \simeq A \cdot (p^2)^{1.54},\ \  G(p^2) \simeq B \cdot (p^2)^{-0.77}$. 
Therefore the existence of non-perturbative power correction does not influence 
the IR solution. 
IR ghost dominance is realized, and the color confinement criterion of Kugo and Ojima is 
satisfied.


\end{document}